\documentclass[aps,showpacs,preprintnumbers,twocolumn,amsmath,amssymb,prb,floatfix]{revtex4} 

\usepackage{graphicx}
\usepackage{color}

\usepackage{ifpdf}

\ifpdf
\usepackage{epstopdf}
\usepackage[pdftex]{hyperref}
\else
\usepackage[hypertex,ps2pdf,dvips,colorlinks,urlcolor=blue,citecolor=blue]{hyperref}
\fi

\newcommand\be{\begin{equation}}
\newcommand\bes{\begin{subequations}}
\newcommand\esu{\end{subequations}}
\newcommand\ee{\end{equation}}
\newcommand\erf[1]        {\eqref{#1}}
\newcommand{\ud}          {\mathrm d}
\newcommand{\e}          {\mathrm e}

\newcommand\mc            {\mathcal}
\newcommand\p             {\partial}

\newcommand\vev[1]{{\langle#1\rangle}}

\newcommand\doi[2]        {\href{http://dx.doi.org/#1}{#2}}

\begin{document}

\title{Analytic results for a quantum quench from free to hard-core one dimensional bosons}

\author{M\'arton Kormos$^{1,2}$, Mario Collura$^1$,  Pasquale Calabrese$^1$}

\affiliation{$^1$Dipartimento di Fisica dell'Universit\`a di Pisa and INFN, 56127 Pisa, Italy\\
$^2$ MTA-BME ÒMomentumÓ Statistical Field Theory Research Group, 1111 Budapest, Budafoki\'ut 8, Hungary }

\date{\today}

\begin{abstract}

It is widely believed that the stationary properties after a quantum quench in  integrable systems 
can be described by a generalized Gibbs ensemble (GGE), even if all the analytical evidence is based on free theories in which the pre- and post-quench modes are linearly related. 
In contrast, we consider the experimentally relevant quench of the 
one-dimensional Bose gas from zero to infinite interaction,  in which the relation between modes is nonlinear, and consequently   
Wick's theorem does not hold. 
We provide exact analytical results for the time evolution of the dynamical density-density 
correlation function at any time after the quench and we prove that its stationary value is described by a 
GGE in which Wick's theorem is restored. 

\end{abstract}


\maketitle

Recent experiments on trapped ultra-cold atomic gases
\cite{uc,kww-06,tc-07,tetal-11,cetal-12,getal-11,shr-12,rsb-13} allowed for the realization and the experimental study of  
(essentially) unitary non-equilibrium evolution on long time scales.  
Among the non-equilibrium situations, the one that attracted most of the attention is the global quantum quench, in which 
the initial state is the ground-state of a translationally invariant Hamiltonian differing from the one governing the evolution  by 
an experimentally tunable parameter such as a magnetic field \cite{revq}.
A key question is whether the system reaches a stationary state, and if it does, 
how to characterize its physical properties at late times without solving the non-equilibrium dynamics. 
It is commonly believed that 
{\it local} observables generally attain a stationary value and, depending on the Hamiltonian governing the time evolution, 
their behavior  either can be described by a thermal 
distribution or by a GGE \cite{gg}, for non-integrable and integrable Hamiltonians, respectively
(see however \cite{kla-07,bch-11,gme-11,gm-11,gp-08} for some criticism). 
Many numerical investigations seem to confirm this scenario \cite{gg,mwn-07,rdo-08,r-09,rs-12,bkl-10,rsm-10,gcg-11,rf-11,bdkm-12,krs-12,sks-13}, 
but due to their intrinsic limitations (such as 
finite size and finite time effects) exact analytic calculations are playing a central role.  
However, while solving the non-equilibrium dynamics of non-integrable models is clearly impossible, even the analytic study of integrable interacting systems in the thermodynamic limit (TDL)
is still beyond our present  capabilities, despite several attempts in this direction \cite{cro,fcc-09,grd-10,fm-10,sfm-12,mc-12,ck-12,a-12,ce-13,bck-13}.
For these reasons, analytic calculations have concentrated on two main aspects.
On the one hand, many studies considered the exact dynamics of models in which both the pre- and post-quench Hamiltonian can be mapped to free particles \cite{cc-06,c-06,cdeo-08,bs-08,CEFII,scc-09,CEF,f-13,eef-12,se-12,fe-13,US,fost}.
%{\red showing that the non-equilibrium dynamics can be difficult to treat analytically even in free models}.
On the other hand, a series of more recent investigations \cite{mc-12b,p-13,fe-13b,ksc-13,m-13,fcce-13} 
attempt to construct the GGE for truly interacting post-quench Hamiltonians starting from 
particular initial states, allowing for numerical or experimental checks of GGE predictions.

However, {\it all} the previous exact analytic studies of the full time-dependence after a quench and the GGE not only considered free theories, but also the 
case in which the pre- and post-quench modes are related by a linear transformation \cite{cc-06,c-06,cdeo-08,bs-08,CEFII,scc-09,CEF,f-13,eef-12,se-12,fe-13,US}, most often a Bogoliubov one (see, however, Ref.\ \onlinecite{fost}). 
In this Letter, we provide the first example in which the GGE works even for a non-linear transformation between modes realized 
in one of the most interesting experimental situations:  the quench from zero to infinite interaction in a one-dimensional Bose gas. 
This quench has been studied in the past \cite{grd-10,ksc-13,fle-10}, but until now resisted {\it an}y analytical computation. 
Apart from the direct interest, our results will also be a benchmark for the novel 
numerically exact methods based on integrability \cite{cro,mc-12,ck-12,a-12,ce-13}.

{\it The model}.
We consider the Lieb--Liniger model, a one-dimensional Bose gas with pairwise delta interaction
on a ring of circumference $L$ with periodic boundary conditions (PBC), i.e. with Hamiltonian \cite{LiebPR130}
\be\label{HLL}\hspace{-2mm}
H = \int_{0}^{L} dx \big[\partial_x \hat\phi^{\dagger} (x) \partial_x \hat\phi (x) 
+ c\, \hat\phi^{\dagger} (x) \hat\phi^{\dagger} (x) \hat\phi (x) \hat\phi(x) \big],
\ee
where $\hat\phi(x)$ is a canonical boson field, $c$ the coupling constant and we set $\hbar=2m=1$.

\begin{figure*}[t]
\includegraphics[width=0.33\textwidth]{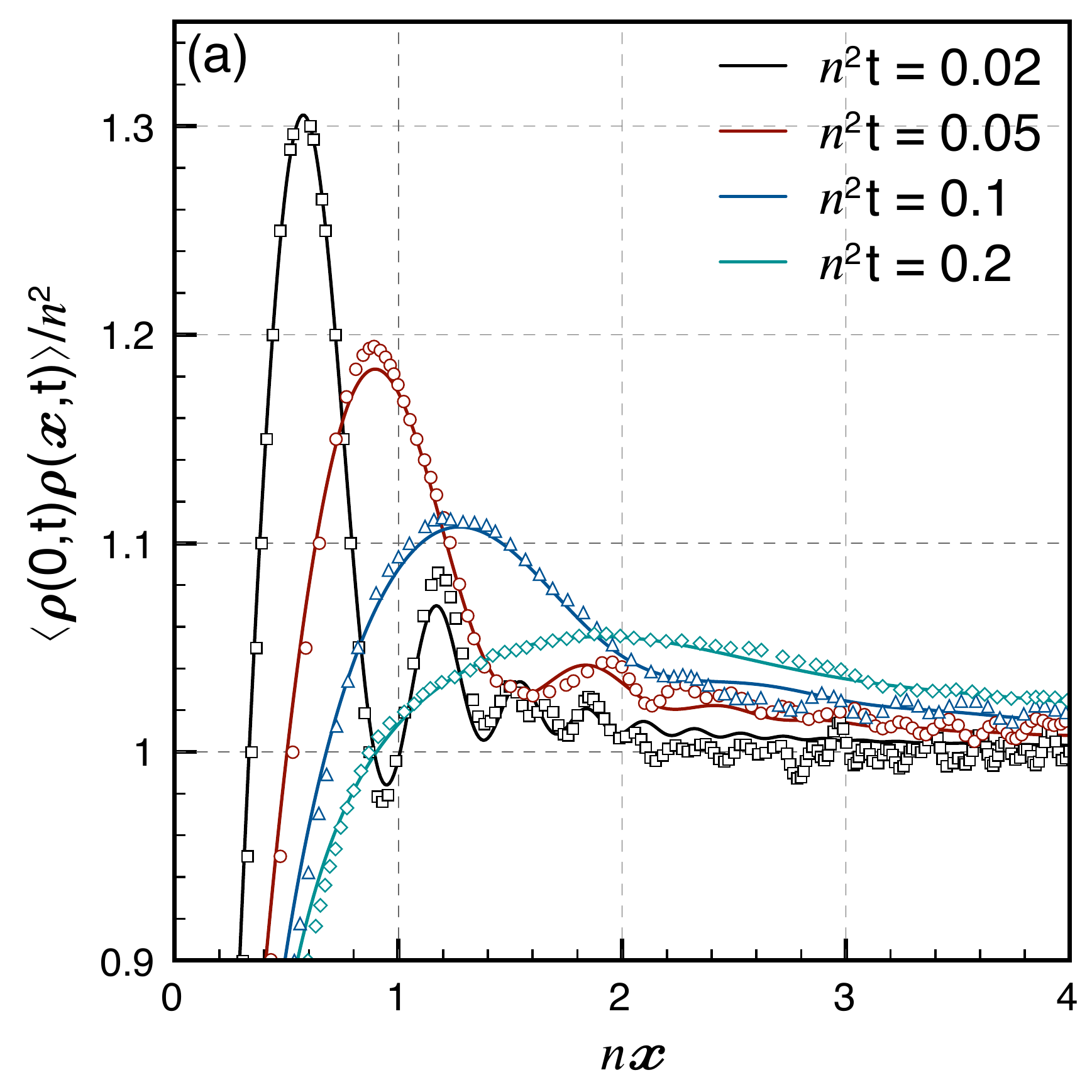}\includegraphics[width=0.33\textwidth]{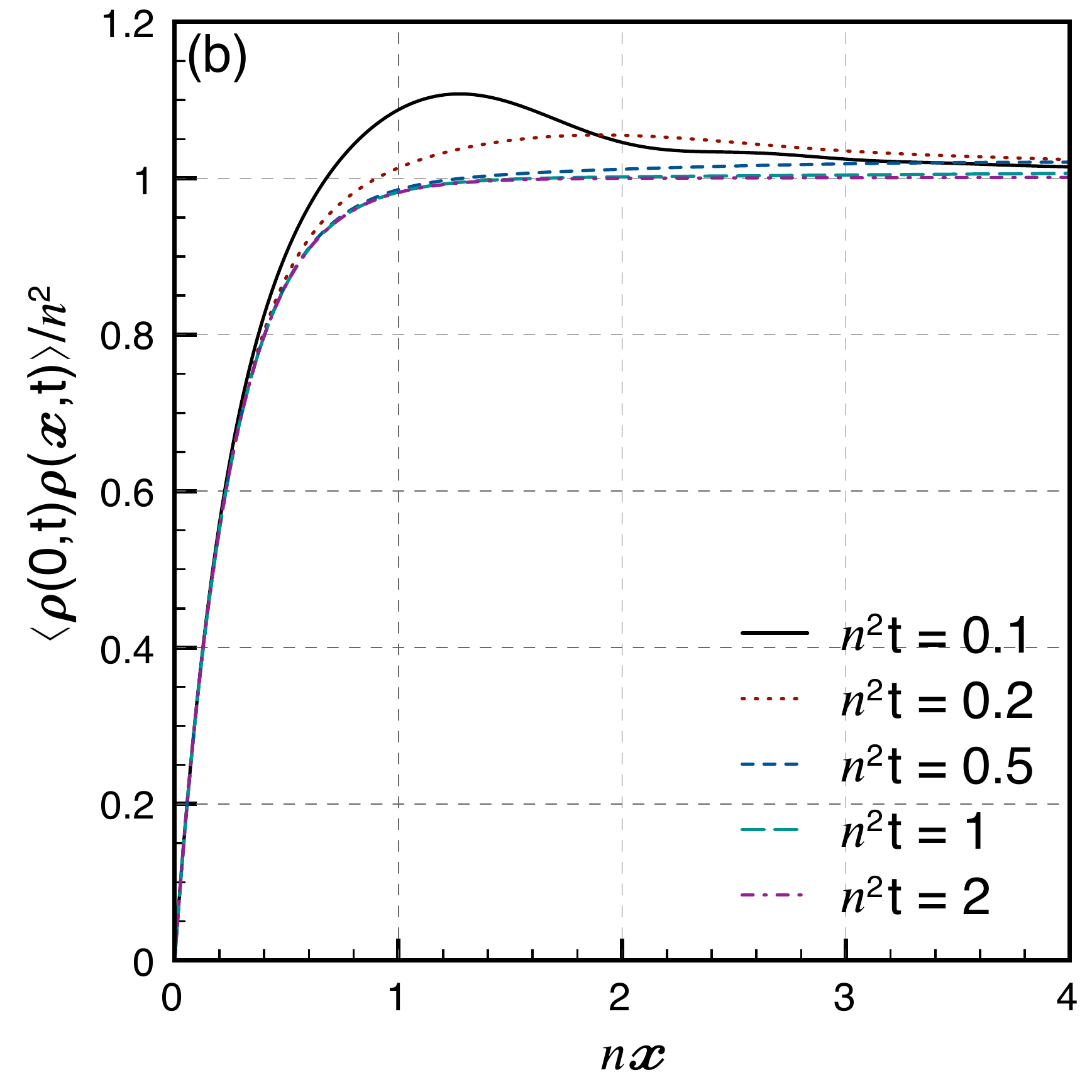}
\includegraphics[width=0.33\textwidth]{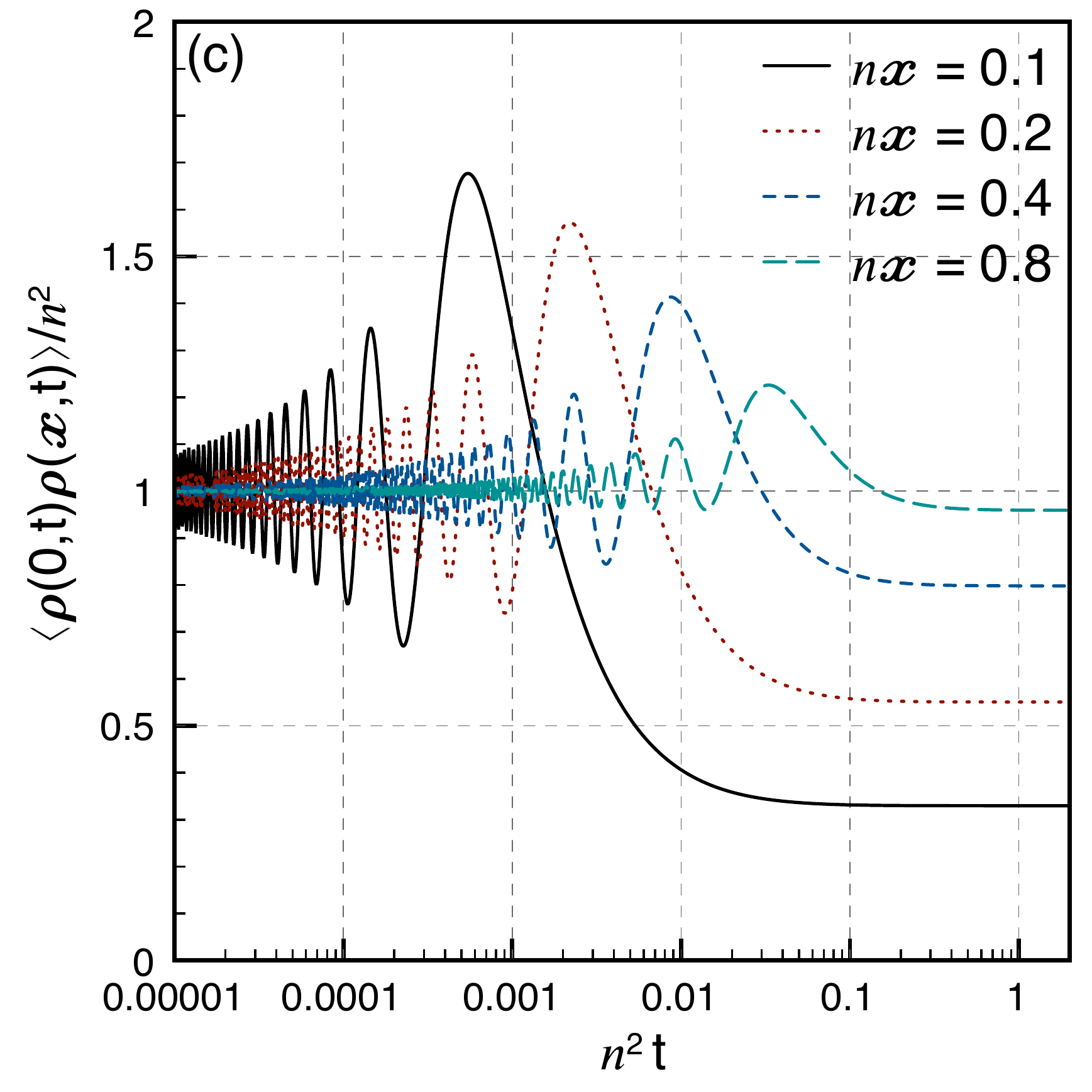}
\caption{Equal-time density-density correlation function $\vev{\hat\rho(0,t)\hat\rho(x,t)}$.
(a) As a function of the distance, we report the correlation for several different times. 
Eq. (\ref{eqtime}) (solid line) is compared with the numerical results of Ref.\ \onlinecite{grd-10} (points)
showing an excellent agreement before the form-factor truncation and finite-size effects become important.
(b) The same as in (a) for larger time.
(c)  $\vev{\hat\rho(0,t)\hat\rho(x,t)}$ for a few fixed $x$ as a function of time in logarithmic scale.
Notice the highly oscillatory (and telescopic in $x$) behavior for very short time that is due to the presence of 
very high energy modes in the initial state.  
After reaching a global maximum, the correlation monotonically reaches the GGE value 
(the large time plateau). 
} 
\label{fig1}
\end{figure*}

We prepare the many-body system in the $N$-particle ground state of the free boson Hamiltonian  
%$H_0$ 
given by Eq. (\ref{HLL}) with $c=0$. 
Writing $\hat{\phi}(x) = \frac{1}{\sqrt{L}}\sum_{q}\mathrm{e}^{i q x} \hat{\xi}_{q}$ where $q=2\pi m/L$ with 
$m$ integer, the ground state is $|\psi_0(N)\rangle=\frac1{\sqrt{N!}}\hat\xi^N_0|0\rangle$.
 We are interested in the TDL, when $N,L\to\infty$ with the particle density $n=N/L$ fixed.
At time $t=0$, we suddenly turn on an infinitely strong interaction, and the evolution 
is governed by the Hamiltonian (\ref{HLL})
with $c=\infty$. It is convenient to rewrite the Hamiltonian in terms of hard-core boson operators, $\hat \Phi,\hat \Phi^\dag$, 
obeying the effective Pauli principle induced by the infinite repulsion  \cite{TG}. 
The constraint that there cannot be two particles at the same point in space is implemented by the algebraic relations 
\be
[\hat\Phi(x)]^2=[\hat\Phi^{\dag }(x)]^2=0\,,\quad \{\hat\Phi(x),\hat\Phi^\dag(x)\}=1\,,
\ee
together with the usual bosonic commutation for $x\neq y$, $[\hat\Phi(x),\hat\Phi(y)]=[\hat\Phi(x),\hat\Phi^\dag(y)]=0$. The Hamiltonian becomes
$H=\int\ud x \,\p_x\hat\Phi^\dag\p_x(x)\hat\Phi(x)$, 
and the commutation relations encode the interactions seemingly absent from the quadratic Hamiltonian.
The {\it non-linear} relation between the pre- and post-quench boson operator can be written as 
$\hat \Phi^{(\dagger)}(x)=P_x \hat \phi^{(\dagger)}(x) P_x$, with $P_x=|0\rangle\langle0|_x+|1\rangle\langle1|_x$ being the local projector on the truncated Hilbert space with at most one boson in $x$.
Strictly speaking our initial state does not belong to the restricted Hilbert space with no local multiple occupation, but
in appendix \ref{App} we show how this problem can be circumvented.

The Jordan--Wigner (JW) transformation
$\hat{\Psi}(x)  =  \exp\left\{i\pi\int_{0}^{x}\ud z\hat{\Phi}^{\dag}(z) \hat{\Phi}(z)\right\}\, \hat{\Phi}(x)$
maps the hard-core boson Hamiltonian onto the  free fermionic one \cite{TG}
\be   
H = \int \ud x \, \p_x\hat{\Psi}^{\dag}(x) \, \p_x \hat{\Psi}(x)\,,
\ee
 diagonalized by the modes $\hat{\eta}_{k}$ and 
$\hat{\eta}^{\dag}_{k}$ (with $k=2\pi m/L$,  $m$ integer or half-integer depending on the parity of $N$):
\be
H= \sum_{k=-\infty}^{\infty} k^2 \hat{\eta}^{\dag}_{k}\hat{\eta}_{k},\quad\hat{\eta}_{k} = \int_{0}^{L}\ud x\,
\frac{\mathrm{e}^{-ikx}}{\sqrt{L}}\hat{\Psi}(x)\,.
\ee

{\it Summary of the results}.
Because of the integrability of the final Hamiltonian (\ref{HLL}), it is expected that the reduced density matrix of 
any finite interval (in the sense described in Refs. \onlinecite{cdeo-08,bs-08,CEFII}) is described by the GGE \cite{gg}
\be
\rho_{GGE}= Z^{-1} e^{-\sum \lambda_i \hat I_i}\,, 
\ee
where $\{\hat I_i\}$ is a complete set of {\it local} integrals of motion and the 
Lagrange multipliers $\lambda_i$ are fixed by the conditions $\langle \psi_0| \hat I_i |\psi_0\rangle={\rm Tr} [\rho_{GGE}\hat I_i]$. 
This GGE has been explicitly constructed for arbitrary final $c$ \cite{ksc-13}.
In the hard-core limit, the final Hamiltonian has a simpler infinite set of conserved charges, 
formed by the fermionic mode occupations, $\hat n(k)$. 
The local conserved charges can be expressed as {\it linear} combinations of the $\hat n(k)$ \cite{dk-90,fe-13,US}, so the GGE's built from $\hat n(k)$ and $\{\hat I_i\}$ are equivalent.

As a first result, we find that the time-independent value of the fermionic mode occupation is \footnote{
The high-momentum tail of $n(k)$ is $\propto k^{-2}$, violating the Tan relation \cite{tan}, but 
this is a manifestation of the diverging energy in the initial state.}
\be
n(k)\equiv\vev{\hat n(k)} = \frac{4n^2}{k^2+4n^2}\,.
\label{nk}
\ee
Clearly also its Fourier transform, the fermionic two-point correlation 
function $\vev{\hat\Psi^\dag(x)\hat\Psi(y)}$, is time independent 
This is not true for the bosonic two-point function $\langle \hat{\phi}^{\dag}(x) \hat{\phi}(y) \rangle$ because 
it contains an infinite string of fermionic operators and the fermionic multi-point functions 
do not factorize into two-point functions. In other words, Wick's theorem does not hold and it  
is restored only for infinite time,  i.e. in the GGE. 
In this case, the bosonic correlation equals the fermionic one,
$\langle \hat{\phi}^{\dag}(x) \hat{\phi}(y) \rangle_{\rm GGE}= n e^{-2n |x-y|}$  (see also Ref. \onlinecite{ksc-13}).
We emphasize that $n(k)$ is the only ingredient needed for the construction of the GGE 
and, via Wick's theorem, it allows for the calculation of any correlation function of local operators, 
showing that the GGE indeed captures the complete stationary behavior.
We stress that, as an important difference \cite{ksc-13} from the local integral of motions $\hat I_i$, the mode occupations $n(k)$ are 
finite in our continuum theory,   so no regularization is necessary.

More complicated and interesting  is the calculation of the (dynamical) density-density correlation function
$\vev{\hat\rho(x_1,t_1)\hat\rho(x_2,t_2)}$ with $\hat\rho(x,t)\equiv \hat{\Psi}^{\dag}(x,t)\hat{\Psi}(x,t)$ 
(%we exploited the fact that, thanks to the properties of the JW string, 
the fermionic density coincides with the hard-core boson density and
the difference between the latter and the true bosonic density vanishes in the TDL, see App. \ref{App}).

We determine the full time-dependence of the dynamical correlation that in the TDL takes the form
\begin{multline}
\vev{\hat\rho(x_1,t_1)\hat\rho(x_2,t_2)} = 
n^2+F_0(\Delta x,\Delta t)F_1(\Delta x,\Delta t)\\
-|F_1(\Delta x,\Delta t)|^2+ |F_2(\Delta x,t_1+t_2)|^2\,,
\label{result}
\end{multline}
where $\Delta x=x_2-x_1$, $\Delta t=t_2-t_1$, and 
\begin{align}
F_0(x,t)&=\int \frac{\ud k}{2\pi} \e^{-i k x+ik^2  t}=\frac{1+\mathrm{sgn}(t)i}{2\sqrt{2\pi|t|}}\e^{-i\frac{x^2}{4t}}\,,\\
F_1(x,t) &= \int \frac{\ud k}{2\pi} \e^{i k x-ik^2  t} n(k)\,, \nonumber \\
F_2(x,t) &= \frac1{2n}\int \frac{\ud k}{2\pi} \e^{i k x+ik^2  t} k n(k)\,. \nonumber
\end{align}
For $x_1=x_2$ we have $F_2(0,t)=0$ because the integrand is an odd function of $k$, thus the auto-correlation function does 
not depend on the time after the quench. 
This is exactly what was observed in the numerical calculation of Ref.~\onlinecite{grd-10}, but remained without explanation until now.

In  the GGE, being diagonal in $\hat n(k)$,  the correlation function is given by Eq.~\erf{result} without the last term, 
so it coincides with the $t\to\infty$ limit. 
This shows that the GGE correctly captures the dynamical correlation function in the large time limit for any $\Delta x, \Delta t$. 
Given that the auto-correlation does not depend on time, the GGE result in this case turns out to be exact at any finite time.

\begin{figure*}[t]
\includegraphics[width=0.48\textwidth]{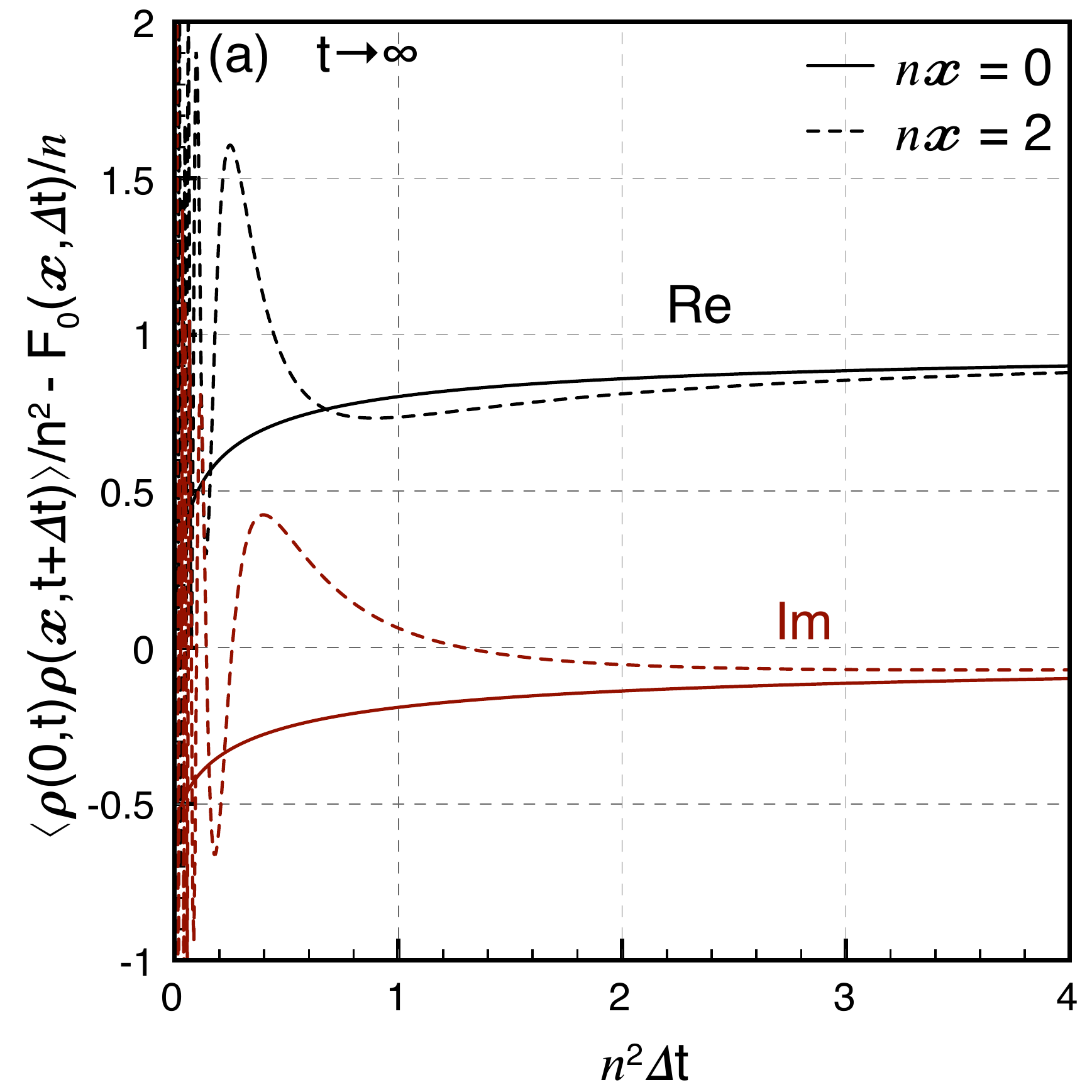}\includegraphics[width=0.48\textwidth]{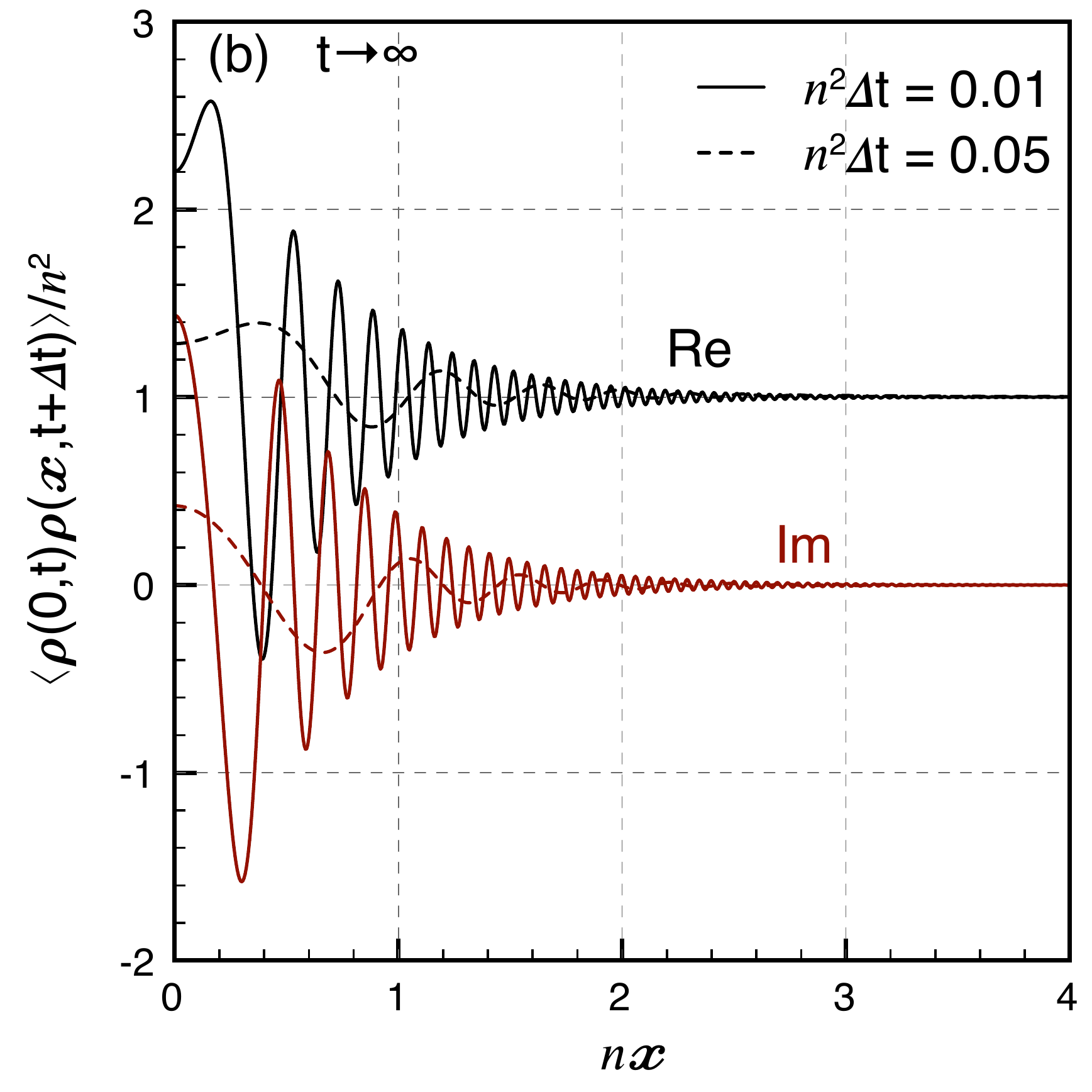}
\caption{Large time dynamical density-density correlation function $\vev{\hat\rho(0,t)\hat\rho(x,t+\Delta t)}$.
(a) Subtracted correlation (i.e. $\vev{\hat\rho(0,t)\hat\rho(x,t+\Delta t)}- n F_0(x,\Delta t)$ to avoid the 
divergence at $x=\Delta t=0$)
as function of $\Delta t$ for $x=0$ and $nx=2$. 
The autocorrelation ($x=0$) does not depend on the elapsed time $t$, so the plot is valid for any $t>0$.
The real part agrees perfectly with the numerical data in Ref. \onlinecite{grd-10}, but the 
imaginary part does not because of a different subtraction. 
(b) Full correlation as function of $x$ for $n^2 \Delta t=0.01$ and $n^2 \Delta t=0.05$.
} 
\label{fig2}
\end{figure*}

The equal-time density-density correlator is included as a special case for $t_1=t_2=t$, for which   we obtain
\begin{multline}
\vev{\hat\rho(x_1,t)\hat\rho(x_2,t)} =
n^2+n\e^{-2n|x_1-x_2|}\delta(x_1-x_2)\\
-n^2\e^{-4n|x_1-x_2|} + |F_2(\Delta x,2t)|^2 \,.
\label{eqtime}
\end{multline}
This result is shown and discussed  in Fig. \ref{fig1}, while the dynamical correlation function is  reported in Fig. \ref{fig2}.
Some qualitative features of these figures resemble the 3D results in Bogoliubov approximation \cite{nm-13}.

For large time we can  define the dynamical structure factor as the double Fourier transform of the connected density-density correlation in $\Delta x$ and $\Delta t.$ 
A straightforward calculation leads to 
\be
S(q,\omega)=\frac{8n^2(q^2+\omega)^2|q|}{[(4nq)^2+(q^2-\omega)^2][(4nq)^2+(q^2+\omega)^2]}.
\ee
This expression satisfies the 
$f$-sum rule $\int d\omega S(q,\omega)\omega=2\pi n q^2$, providing a non-trivial test for our results.

{\it Fermionic occupation numbers}.
We calculate $\vev{\hat n(k)}$ in the initial state through its Fourier transform
$\vev{\hat\Psi^\dag(x)\hat\Psi(y)}$. 
We rewrite the fermionic operators in terms of the hard-core boson  using the inverse of the JW mapping. 
The two-point function takes the form
\begin{multline}
\label{2point}
\langle \hat{\Psi}^{\dag}(x) \hat{\Psi}(y) \rangle = \sum_{j=0}^{\infty}\frac{(-2)^j}{j!}\int_{x}^{y}dz_1\cdots\int_{x}^{y}dz_j\,\\
\langle \hat{\Phi}^{\dag}(x) \hat{\Phi}^{\dag}(z_1)\cdots\hat{\Phi}^{\dag}(z_j) \hat{\Phi}(z_j)\cdots\hat{\Phi}(z_1)
\hat{\Phi}(y)\rangle\,,
\end{multline}
where the factor $(-2)^j$ results from normal ordering. 
We proceed by treating the hard-core boson fields as they were canonical bosonic fields. The validity of this approach is fully analyzed and is derived from a complete rigorous 
lattice calculation in \ref{App}. Using  $\hat{\xi}_{q}|\psi_0(N)\rangle = \delta_{q0}\sqrt{N} |\psi_0(N-1)\rangle$, one obtains
\begin{multline}
\langle \hat{\phi}^{\dag}(x) \hat{\phi}^{\dag}(z_1)\cdots\hat{\phi}^{\dag}(z_j) \hat{\phi}(z_j)\cdots\hat{\phi}(z_1)
\hat{\phi}(y)\rangle = \\
\frac{1}{L^{j+1}} \langle N | (\hat{\xi}^{\dag}_{0})^{j+1}(\hat{\xi}_{0})^{j+1}| N \rangle = \frac1{L^{j+1}} \frac{N!}{(N-j-1)!}\,.
\end{multline}
Finally, integrating over $z_1,\dots,z_j$, we have
\begin{multline}
\langle \hat{\Psi}^{\dag}(x) \hat{\Psi}(y) \rangle  = \frac{N}L \sum_{j=0}^{\infty}\frac{[-2|x-y|/L]^j}{j!}\frac{(N-1)!}{(N-j-1)!}\\
 =  n \left(1- \frac{2n|x-y|}{N}\ \right)^{N-1} \xrightarrow[]{N\to\infty} n \mathrm{e}^{-2n|x-y|}\,.
\end{multline}
The momentum distribution function is obtained by Fourier transformation leading to Eq. (\ref{nk})

{\it The dynamical density-density correlation function} is
\begin{multline} \hspace{-4mm}
\vev{\hat\rho(x_1,t_1)\hat\rho(x_2,t_2)} = 
\frac1{L^2}\sum_{k_1,k_2,k_3,k_4}\e^{-i(k_1-k_2)x_1-i(k_3-k_4)x_2}\\
\e^{i(k_1^2-k_2^2)t_1}\e^{i(k_3^2-k_4^2)t_2}
\langle \psi_0|\hat \eta^\dag_{k_1}\hat\eta_{k_2}\hat\eta^\dag_{k_3}\hat\eta_{k_4}|\psi_0\rangle\,,
\end{multline}
and so we need to evaluate the initial fermionic four-point  correlation 
\begin{eqnarray}\label{4point}
 && \langle \psi_0|
 \hat{\eta}^{\dag}_{k_1}  \hat{\eta}_{k_2}  \hat{\eta}^{\dag}_{k_3}  \hat{\eta}_{k_4}|\psi_0\rangle=
\frac1{L^2}\int_{0}^{L}\ud z_1\ud z_2\ud z_3\ud z_4 \\
 && \hspace{-4mm} \e^{i(k_1z_1-k_2z_2+k_3z_3-k_4z_4)}
\langle \psi_0|\hat \Psi^\dag(z_1)\hat\Psi(z_2)\hat\Psi^\dag(z_3)\hat\Psi(z_4)|\psi_0\rangle.
\nonumber
\end{eqnarray}
The four-point function 
$\langle \psi_0|\hat \Psi^\dag(z_1)\hat\Psi(z_2)\hat\Psi^\dag(z_3)\hat\Psi(z_4)|\psi_0\rangle$
can be calculated analogously to the two-point function. 
Let us first consider the case $z_1<z_2 < z_3<z_4$, with two JW strings, one between $z_1$ and $z_2$ 
and one between $z_3$ and $z_4$. Operators belonging to different strings commute so it is easy to normal order them. 
The expectation value is then 
\be
\vev{ \hat{\Psi}^{\dag}(z_1) \hat{\Psi}(z_2)\hat{\Psi}^{\dag}(z_3) \hat{\Psi}(z_4)}%_{z_1<z_2 < z_3<z_4}=\\
=n^2 \mathrm{e}^{-2n(z_4-z_3+z_2-z_1)} .
\ee
If $z_1<z_2 < z_3<z_4$ does not hold, but  the $z_i$ are all distinct,
one needs to reorder the operators. 
In other words, we break up the domain of the four-dimensional integral in Eq. (\ref{4point}) 
into regions defined by the order of $z_i$. 
In each of these regions $z_{\mc{P}_1}< z_{\mc{P}_2}< z_{\mc{P}_3}<z_{\mc{P}_4}$, 
where $\mc{P}$ is one of the 24 permutations of the numbers $\{1,2,3,4\}$.
While reordering the operators, we pick up signs due to their fermionic nature.
There are also extra minus signs coming from commuting the operators and the JW strings. For each permutation the result is 
$\sigma_{\mc{P}} \,n^2 \e^{-2n(z_{\mc{P}_4}-z_{\mc{P}_3}+z_{\mc{P}_2}-z_{\mc{P}_1})}$, 
where $\sigma_{\mc{P}}$ is the overall sign in the permutation $\mc{P}.$ Finally, one needs to deal with the cases when two or more of the four operators are at the same point, which leads to a contact term  $\delta(z_2-z_3)n \e^{-2n|z_4-z_1|}$.
In summary, the four-point function is given by 
\begin{multline}
\langle \hat{\Psi}^{\dag}(z_1) \hat{\Psi}(z_2)\hat{\Psi}^{\dag}(z_3) \hat{\Psi}(z_4) \rangle=
\delta(z_2-z_3)n \e^{-2n|z_4-z_1|}\\
+\sum_{\mc{P}}\theta(z_{\mc{P}})\sigma_\mc{P} \,n^2\e^{-2n(z_{\mc{P}_4}-z_{\mc{P}_3}+z_{\mc{P}_2}-z_{\mc{P}_1})}\,,
\end{multline}
where  $\theta(z_{\mc{P}})=\theta(z_{\mc{P}_4}-z_{\mc{P}_3})\theta(z_{\mc{P}_3}-z_{\mc{P}_2})\theta(z_{\mc{P}_2}-z_{\mc{P}_1}).$

For the momentum space four-point function (\ref{4point}), we need to evaluate the four-dimensional integral 
over $z_1,\dots, z_4$. The integral of the first term gives
\be\hspace{-2mm}
\frac1{L^2}\int_{0}^{L} \ud^4 z\,\delta(z_{23})\,n \e^{-2n|z_{14}|} = 
\delta_{k_2,k_3}\delta_{k_1,k_4} \frac{4n^2}{4n^2+k_1^2},
\ee
where $z_{ij}=z_i-z_j.$ The sum over permutation leads to the sum of $24$ spatially ordered integrals. The first permutation gives
%
%\be
%\begin{split}
\begin{multline}
I_{\mathcal{\mc{P}}_1} =  \frac{n^2}{L^2}\int_{\mc{P}_1} \ud^4 z \,
\e^{i[k_1 z_1 - k_2 z_2 + k_3 z_3 -k_4z_4]} \e^{-2n(z_{43}-z_{21})}\\
 = \frac{n^2}{2} \frac{\delta_{k_1,k_2}\delta_{k_3,k_4}}{(ik_1+2n)(ik_3+2n)}+\dots\,,
\end{multline}
%\end{split}
%\ee
where $\int_{\mc{P}_1} \ud^4 z=\int_{0<z_1<z_2<z_3<z_4<L} \ud^4 z$, and the ellipses stand for terms subleading in $L$. The integral of a permutation on the variables $z_j$ can be translated to a permutation in the momentum variables $k_j$, and we finally obtain
\begin{multline}
\langle \psi_0 | \hat{\eta}^{\dag}_{k_1}  \hat{\eta}_{k_2}  \hat{\eta}^{\dag}_{k_3}  \hat{\eta}_{k_4}|\psi_0 \rangle =
n(k_1)\delta_{k_2,k_3}\delta_{k_1,k_4}\\
+(\delta_{k_1,k_2}\delta_{k_3,k_4}-\delta_{k_2,k_3}\delta_{k_1,k_4})n(k_1)n(k_3) \\
+ \delta_{k_1,-k_3}\delta_{k_2,-k_4}\frac{ k_1 k_2}{4n^2}n(k_1)n(k_2)\,.
\label{4pt}
\end{multline}
 The result coincides with the correlation in a state satisfying $\vev{\hat\eta^\dag_k\hat\eta_l}=\delta_{k,l}n(k)$, 
 apart from the last term showing that the initial state is a superposition of free-fermion states consisting of pairs of particles with opposite momenta \cite{grd-10,kpz}. 
%Only this last term contributes to the time-dependence of the correlator.

Taking the limit $L\to\infty$, the momentum sums become integrals and the final result is Eq. (\ref{result}).
In this equation, the first term, $n^2$, comes from $\delta_{k_1,k_2}$ in Eq.~\erf{4pt}, 
the second comes from the contact term proportional to $\delta_{k_2,k_3}$ while the third term from the 
$\delta_{k_2,k_3}$ in the second line of Eq. (\ref{4pt}). 
The only term depending on $t_1$, the time after the quench, is the fourth one which originates from the anomalous 
$\delta_{k_1,-k_3}$ contribution in Eq.~\erf{4pt}.

{\it Conclusions}.
We provided analytic results for the dynamical density-density correlation in a one-dimensional Bose gas
after a quench from  free to hard-core bosons, which agree with previous numerical investigations \cite{grd-10}. 
Although this is a quench between two free theories, the pre- and post-quench mode-operators are not linearly related 
and so the multipoint correlations in momentum space {\it do not factorize} in terms of two-point ones, preventing us from the 
use of standard techniques. The factorization is recovered only for infinite time when, it turns out, the GGE completely 
describes the system.
Conversely, for finite times, the multipoint correlations of the modes must be calculated explicitly making very laborious 
the determination of the time-dependence of other correlations, such as the bosonic two-point function,
which, contrarily to other cases in the literature, cannot be expressed as a Fredholm minor.

{\it Acknowledgments}.   
All authors acknowledge the ERC  for financial  support under  Starting Grant 279391 EDEQS.

\appendix

\section{Lattice formulation of the BEC to TG quench}
\label{App}

\subsection{Setup on the lattice}

 Let us consider a system of $N$ bosons hopping on a one-dimensional lattice composed of $M$ sites, 
with lattice spacing $\delta$.
As in the main text, we want to describe the quench from free bosons to hard-core ones. 
The initial state is then the %free lattice boson ground state or 
BEC:
\be
|\text{BEC}\rangle_N = \frac1{\sqrt{M^N N!}}\left(\sum_{i=1}^M b_i^\dag \right)^N|0\rangle\,,
\label{BEClat}
\ee
where the $b_i^{(\dag)}$ are canonical boson operators and $|0\rangle=\prod_{\otimes i}|0\rangle_i$ with $|n\rangle_i$ being the $n$-boson state at site $i$. 

Let us introduce hard-core boson operators $a_i$ that satisfy the algebra
\begin{align}
[a_i,a_j]=[a_i^\dag,a_j^\dag]=[a_i,a_j^\dag]&=0 \qquad i\neq j\,,\\
a_i^2=a_i^{\dag 2}=0\,,\quad \{a_i,a_i^\dag\}&=1\,.
\end{align}
The relation between the canonical and hard-core boson operators, quite obviously, is
\be
a_i=P_ib_iP_i\,,\qquad
a_i^{\dag}=P_ib_i^{\dag}P_i\,,
\ee
where $P_i=|0\rangle\langle0|_i+|1\rangle\langle1|_i$ is the on-site projector on the truncated Hilbert space. 

We can map  hard-core bosons to free fermions through the Jordan--Wigner transformation
\begin{subequations}
\begin{align}
a_i&=e^{-i\pi\sum_{j<i}c^\dag_jc_j}c_i=\prod_{j<i}(1-2c_j^\dag c_j)\,c_i\,,\\
c_i&=e^{i\pi\sum_{j<i}a^\dag_ja_j}a_i=\prod_{j<i}(1-2a_j^\dag a_j)\,a_i\,.
\end{align}
\end{subequations}
 The lattice thermodynamic limit is obtained as $N, M \to\infty$ keeping the filling fraction $\nu = N/M$ constant.
However here we will be also interested  in the continuum limit in finite systems, i.e. we let the lattice spacing 
$\delta\to0$, the number of sites $M \to \infty$ while physical length $L$ is kept constant $L=M\delta$.
The continuum TDL can be now taken as $N,L\to\infty$, with the gas density $n=N/L$ constant.
We shall see that the order of these two limits, while it could be important in general, does not matter 
for the observables we are interested in.
Finally, to have a fully defined continuum limit, we need the following relations between lattice operators and continuum ones 
\be
b_m=\sqrt{\delta}\,\hat\phi(m\delta)\,,\quad a_m=\sqrt{\delta}\,\hat\Phi(m\delta)\,,\quad
c_m=\sqrt{\delta}\,\hat\Psi(m\delta)\,.
\ee

\subsection{Fermionic momentum distribution}

The real space correlation function of $c_j$ operators reads (for $k<l$)
\begin{multline}
_N\langle\text{BEC}|c_k^\dag c_l |\text{BEC}\rangle_N=\\
_N\!\langle\text{BEC}|a_k^\dag \prod_{j=k}^{l-1}(1-2a_j^\dag a_j)\,a_l |\text{BEC}\rangle_N=\\
_N\!\langle\text{BEC}|a_k^\dag \sum_{r=0}(-2)^r  
\hspace{-5mm} \sum_{k<n_1<n_2<\dots n_r<l} 
\hspace{-8mm}  a_{n_1}^\dag a_{n_1}\dots a_{n_r}^\dag a_{n_r} a_l |\text{BEC}\rangle_N,
\end{multline}
where we used the fact that due to the hardcore condition the product effectively runs from $j=k+1$. Each term in the sum has the form
\be
_N\langle\text{BEC}|a_k^\dag (a_{n_1}^\dag a_{n_1}\dots a_{n_r}^\dag a_{n_r} )a_l |\text{BEC}\rangle_N\,.
\ee
In order to calculate this expectation value let us start by expanding 
the term $\left(\sum_{i=1}^M b_i^\dag \right)^N$ in the BEC state (\ref{BEClat})
\be
|\text{BEC}\rangle_N = \frac1{\sqrt{M^N N!}}
\sum_{i_1,\dots,i_M}\binom{N}{i_1,\dots,i_M} b_1^{\dag i_1}\dots b_M^{\dag i_M}|0\rangle,
\ee
where the sum runs over all sets of non-negative integers $\{ i_1,\dots,i_M\}$ such that their sum $\sum_j i_j=N$.
The string of $a$ operators can be replaced by a string of canonical $b$ operators if we insert the projectors $P_i$. Starting from the right we have $a_l=P_l b_l P_l$. The rightmost $P_l$ projects out all terms in the multinomial expansion that have more than one particle at site $l$. However, if there is no particle at site $l$, the $b_l$ operator annihilates the state, thus we find that $i_l=1$ in order to have a non-zero result. The second $P_l$ can be dropped because now there is no particle at site $l$. Moving to the next operator, we 
similarly find that $i_{n_r}=1$ must hold and $a_{n_r}^\dag a_{n_r}$ simply takes the eigenvalue $1$. 
Continuing all the way to the left we find
\begin{widetext}
\begin{multline}
a_k^\dag (a_{n_1}^\dag a_{n_1}\dots a_{n_r}^\dag a_{n_r} )a_l |\text{BEC}\rangle_N=\\
 \frac1{\sqrt{M^N N!}} \sum_{\{i_1,\dots,i_M\}'}
 \binom{N}{i_1,\dots,i_k=0,\dots,i_{n_1}=1,\dots,i_l=1,\dots,i_M} b_1^{\dag i_1}\dots b_M^{\dag i_M}|0\rangle\,,
\end{multline}
\end{widetext}
where $\{i_1,\dots,i_M\}'=\{i_1,\dots,i_M\}\backslash \{i_k,i_{n_1},\dots,i_{n_r},i_l\}$ and all $b_{n_j}^\dag$  and $b_k^\dag$
 come with power one, while there is no $b_l^\dag$.

Now we have to take the scalar product with the BEC bra state. 
Clearly, the only non-zero contributions come from the products of monomials of the bra and the ket state in which the powers of $b$ operators match perfectly.  Using $\langle0|b_i^nb_i^{\dag n}|0\rangle=n!$ this leaves us with
\begin{multline}
_N\langle\text{BEC}|a_k^\dag (a_{n_1}^\dag a_{n_1}\dots a_{n_r}^\dag a_{n_r} )a_l |\text{BEC}\rangle_N= \\
\frac1{M^N N!}\sum_{\{\bar i_1,\dots,\bar i_{M-r-2}\}}\left(\frac{N!}{\bar i_1!\cdots \bar i_{M-r-2}!}\right)^2\bar i_1!\cdots\bar i_{M-r-2}!\,,
\label{eq}
\end{multline}
where $\{\bar i_1,\dots,\bar i_{M-r-2}\}=\{i_1,\dots,i_M\}'$ by relabelling. 
Note that $\sum_{j=1}^{M-r-2}\bar i_j = N-r-1$.
So, rewriting the previous formula as 
\begin{multline}
_N\langle\text{BEC}|a_k^\dag (a_{n_1}^\dag a_{n_1}\dots a_{n_r}^\dag a_{n_r} )a_l |\text{BEC}\rangle_N
=\\ \frac1{M^N}N(N-1)\cdots(N-r)\sum_{\{\bar i_1,\dots,\bar i_{M-r-2}\}}\frac{(N-r-1)!}{\bar i_1!\cdots\bar i_{M-r-2}!}\,,
\end{multline}
we can use the expression for the sum of multinomial coefficients and we arrive at
\begin{multline}
_N\langle\text{BEC}|a_k^\dag (a_{n_1}^\dag a_{n_1}\dots a_{n_r}^\dag a_{n_r} )a_l |\text{BEC}\rangle_N =\\ 
\frac1{M^N}N(N-1)\cdots(N-r)\cdot (M-r-2)^{N-r-1}=\\
\frac{N}M \frac{N-1}M\cdots\frac{N-r}M\cdot\left(1-\frac{r+2}M\right)^{N-r-1}\,.
\label{locproj}
\end{multline}

Note that the case of $_N\langle\text{BEC}|a_l^\dag a_l |\text{BEC}\rangle_N$ is a bit different. 
The only condition we get is $i_l=1$, so by the previous logic we obtain
\begin{multline}
_N\langle\text{BEC}|a_l^\dag a_l |\text{BEC}\rangle_N = \\
\frac1{M^N}N\sum_{\{\bar i_1,\dots,\bar i_{M-1}\}}\frac{(N-1)!}{\bar i_1!\cdots\bar i_{M-1}!}=\\
\frac{N}{M^N} (M-1)^{N-1}=\nu\left(1-\frac1M\right)^{N-1}\approx\nu e^{-\nu}\,,
\end{multline}
where in the last step we took the thermodynamic limit on the lattice. 
Note that $\langle a_l^\dag a_l\rangle\neq\nu$, which is the manifestation of the fact that the operator $a_l^\dag a_l$ is not the 
physical particle density operator, but it is only on the restricted Hilbert space, while the BEC state lies outside of this subspace. However, in the continuum limit, $\nu=n\delta\to0$, the difference vanishes.

Going back to the case of general $N$ and $M$ we can calculate now the fermonic correlation function. We take the continuum limit before the thermodynamic one. In this case the maximal length of the string is $N-1$, because the string contains number operators at different sites and it acts on a $N-1$ particle state. So we have
\begin{widetext}
\begin{multline}
_N\langle\text{BEC}|c_k^\dag c_l |\text{BEC}\rangle_N=
\sum_{r=0}^{\mathbf{N-1}}(-2)^r\binom{l-k-1}{r}\frac{N}M \frac{N-1}M\cdots\frac{N-r}M\cdot\left(1-\frac{r+2}M\right)^{N-r-1}\\
%=n \left(1-2\frac{N-1}N\,nx+2\frac{(N-1)(N-2)}{N^2}(nx)^2-\frac43\frac{(N-1)(N-2)(N-3)}{N^3}(nx)^3+\right.\\
%\left.\frac23\frac{(N-1)(N-2)(N-3)(N-4)}{N^4}(nx)^4+\dots\right)=\\
=\delta\cdot n 
\left(1-2\frac{\binom{N-1}{1}}N (nx)+4\frac{\binom{N-1}{2}}{N^2}(nx)^2-8\frac{\binom{N-1}{3}}{N^3}(nx)^3+16\frac{\binom{N-1}{4}}{N^4}(nx)^4+\dots\right)
\,,
\end{multline}
\end{widetext}
where we evaluated the sum explicitly with various finite $N$ and fixed continuum densities $n=N/L$. 
The pattern is quite clear, and  in the thermodynamic limit $N\to\infty$ we obtain
\be
_N\langle\text{BEC}|c_k^\dag c_l |\text{BEC}\rangle_N \longrightarrow \delta \cdot ne^{-2n(x-y)}\,. 
\ee
One gets the same result by taking first the thermodynamic limit on the lattice keeping $\nu$ fixed, and 
taking the continuum limit $\nu\to 0$ as the last step.

Clearly, if $l<k$ nothing changes in the calculation, except that the distance between the two points (the length of the string) is given by $k-l-1$. So the final result, irrespective of the order in which the continuum and thermodynamic limits are taken, is
\be
\langle\text{BEC}|\hat\Psi^\dag (x)\hat\Psi(y) |\text{BEC}\rangle = n e^{-2n|x-y|}\,,
\ee
 which coincides with the result found directly in the continuum limit in the main text.

\subsection{Projected initial state}

We show that the {\it same} result is obtained in the continuum limit if instead of applying the local projectors to the operators, 
one projects out the multiply occupied states from the BEC superposition at the start. 
So we apply a global projector $P=P_1\otimes P_2\cdots P_M$ to the initial state which eliminates all terms from the expansion of the BEC that contain multiple occupancies. What is left is a completely symmetric state of the form $\sum_{j_1,\dots,j_N=1}^Mb^\dag_{j_1}\cdots b^\dag_{j_N}$ where $j_k\neq j_l$. 

Eq.\ \erf{eq}  holds also in this case, since obtaining it we only fixed occupancies on some of the sites based on the operators appearing in the correlation function. The difference is that we now restrict the sum over $i_j$ such that each $i_j$ is at most one. The problem is now to count how many ways we can put $N-r-1$ factors 1 on the remaining $M-r-2$ sites. 
The answer is clearly $\binom{M-r-2}{N-r-1}$, so
\begin{multline}
_N\langle\text{BEC}|a_k^\dag (a_{n_1}^\dag a_{n_1}\dots a_{n_r}^\dag a_{n_r} )a_l |\text{BEC}\rangle_N
=\\ 
\frac1{M^N}N!\frac{(M-r-2)!}{(N-r-1)!(M-N-1)!}\\=
\frac{N}M \frac{N-1}M\cdots\frac{N-r}M\cdot\left(1-\frac{r+2}M\right)\cdots\left(1-\frac{N}M\right)\,.
\end{multline}
This is different from the result of the local projector approach, Eq.~\erf{locproj}, but the difference vanishes in the continuum limit 
when $\nu=N/M\to0$. 
{\it This shows that in the continuum limit the initial BEC state behaves as if it belonged to the hard-core Hilbert space.}

\subsection{Fermionic real-space four-point function in the initial state}

In the calculation of the dynamical density-density correlation function the main ingredient is the real-space fermionic four-point function. Here we compute the corresponding four-point function on the lattice using the same technique as for the two-point function. Let us start with
\be
_N\langle\text{BEC}|c_k^\dag c_l c^\dag _m c_j |\text{BEC}\rangle_N\,,
\label{4corrL}
\ee
where the operators are at different sites in the  spatial order $k<l<m<j$. Rewriting it in terms of hard-core boson operators one gets
\begin{widetext}
\begin{multline}
_N\langle\text{BEC}|c_k^\dag c_l c^\dag _m c_j |\text{BEC}\rangle_N=  
\,_
N\langle\text{BEC}|a_k^\dag \prod_{p=k}^{l-1}(1-2a_p^\dag a_p)\,a_l a_m^\dag \prod_{q=l}^{m-1}(1-2a_q^\dag a_q)\,a_j|\text{BEC}\rangle_N
=\\
=\sum_{r=0}(-2)^r\sum_{s=0}(-2)^s\,
_N\langle\text{BEC}|a_k^\dag\sum_{k<n_1<\dots n_r<l} a_{n_1}^\dag a_{n_1}\dots a_{n_r}^\dag a_{n_r} a_l 
a_m^\dag\sum_{m<m_1<\dots m_s<j} a_{m_1}^\dag a_{m_1}\dots a_{m_s}^\dag a_{m_s} a_j |\text{BEC}\rangle_N\,.
\end{multline}
\end{widetext}
%\ee
Like before, we used the hard-core condition  forcing all operators to be  different in each monomial. 
Now it is just a matter of counting operators and particles. From the expansions of both bra and ket BEC states we must again pick the same term, in which there must be particles only at $r+s+2$ sites labelled by $n_1,\dots,n_r,l,m_1,\dots,m_s,j$, 
while there must be no particle at sites $k,m$. 
Thus we have to distribute $N-(r+s+2)$ particles over the remaining $M-(r+s+4)$ sites. This gives 
\begin{widetext}
\begin{multline}
_N\langle\text{BEC}|c_k^\dag c_l c^\dag _m c_j |\text{BEC}\rangle_N=\\
=\sum_{r=0}(-2)^r\sum_{s=0}(-2)^s\binom{l-k-1}{r}\binom{j-m-1}{s}
%\sum_{k<n_1<\dots n_r<l} \sum_{m<m_1<\dots m_s<j}
\frac{N}M \frac{N-1}M\cdots\frac{N-(r+s+1)}M\cdot\left(1-\frac{r+s+4}M\right)^{N-(r+s+2)}\,.
\end{multline}
\end{widetext}
The double sum runs over $\{r,s\}$ such that $r\le l-k-1$,  $s\le j-m-1$ and $r+s\le N-2$. However, these constraints are imposed automatically! The first two are ensured by the binomial coefficients, and the third one by the product of fractions: if $r+s\ge N-1$ there will be a term with numerator exactly zero. Thus one can set the upper limits formally to $\infty$. In the continuum limit, as $\delta\to0$, we find
\begin{widetext}
\begin{multline}
_N\langle\text{BEC}|c_k^\dag c_l c^\dag _m c_j |\text{BEC}\rangle_N=\\
(\delta\cdot n )^2
\left(\frac{\binom{N-1}{1}}N-4\frac{\binom{N-1}{2}}{N^2} n(x+y)+12\frac{\binom{N-1}{3}}{N^3}n^2(x+y)^2-32\frac{\binom{N-1}{4}}{N^4}n^3(x+y)^3+\dots\right)\,,
\end{multline}
\end{widetext}
where $x=(l-k)\delta$ and $y=(j-m)\delta$, i.e.  the distances between the operators in the two pairs in the continuum limit.
Finally,  in the thermodynamic limit $N\to\infty$, the four-point function becomes
\be
_N\langle\text{BEC}|c_k^\dag c_l c^\dag _m c_j |\text{BEC}\rangle_N\to
\delta^2\cdot n^2 e^{-2n (x+y) }\,.
\ee

If the four operators are not in the order $k<l<m<j$, but they are still at different sites, 
we permute them so that they are spatially ordered. Since they are fermionic operators, we pick up a sign corresponding to 
the order of the permutation.  
Note that the final order of the two creation and the two annihilation operators may be different from the original one. 
Because they are now spatially ordered, we put a string between the first two and the last two. 
However, if any of the strings starts at an annihilation operator one has to include  
in the string product also the first term (i.e. the one $p=k$ below, which was previously discarded), obtaining
\begin{widetext}
\be
c_kc^{(\dag)}_l=a_k \prod_{p=k}^{l-1}(1-2a_p^\dag a_p)\,a^{(\dag)}_l=a_k(1-2a_k^\dag a_k)\prod_{p=j+1}^{l-1}(1-2a_p^\dag a_p)\,a^{(\dag)}_l=-a_k \prod_{p=k+1}^{l-1}(1-2a_p^\dag a_p)\,a^{(\dag)}_l\,,
\ee
\end{widetext}
where we used the hard-core boson algebra.
Now the product is taken over sites between the two operators, like before, but there is an extra minus sign. 
Apart from this sign, the expression is identical to what we had before 
because, although the order of the creation/annihilation operators at the edges is different, 
we are free to rearrange them due to the bosonic commutation of the $a$-operators at different sites. 
The same  is true for a four-point function of the form $c_k^\dag c_m^\dag c_j c_l$, for example: after accounting for 
the minus sign coming from the second pair $c_jc_l$, we are free to rearrange the two creation and two annihilation operators 
outside the strings, restoring the former situation.

So, the recipe to compute $\langle c_k^\dag c_l c^\dag _m c_j \rangle$ is the following. First one should rearrange the operators according to their spatial order, keeping track of the fermionic minus signs. Then one has to multiply by $(-1)^\omega$ where $\omega=0,1,\text{or }2$ is the number of strings starting at an annihilation operator (i.e., the number of annihilation operators at position 1 and 3 in the new order). Then in the continuum and thermodynamic limit, apart from the signs, we have $n^2 e^{-2n(x+y)}$ with $x$ and $y$ being the distances between the first two and the second two operators, respectively.

Finally, we have to consider the cases when two or more operators are at the same site. It turns out that these are continuously connected to the result above as the coordinates approach each other, except when $l\to m$ ($z_2\to z_3$). In this case there is a diverging contribution coming from the commutation relation used to normal order the operators. For example,
\begin{multline}
\delta^{-2}\,\!_N\langle\text{BEC}|c_k^\dag c_l c^\dag _l c_j |\text{BEC}\rangle_N = \\
-\delta^{-2}\,\!_N\langle\text{BEC}|c_k^\dag c^\dag_l c_l c_j |\text{BEC}\rangle_N+ 
\delta^{-2}\,\!_N\langle\text{BEC}|c_k^\dag c_j |\text{BEC}\rangle_N \\
= -n^2 e^{-2n|x|}+\delta^{-1}n e^{-2n|x|}\,,
\end{multline}
where $|x|=|(k-j)\delta|$. No new behavior appears when three or four operators are at the same site or when the four operators are distributed on two sites: these can be obtained as the limits of the formula above. In the continuum limit this contact term gives rise to a $\delta(z_2-z_3)$ contribution. % which needs to be added to the result.

%%%%%%%%%%%%%%%%%%%%%%%%%%%%%%%%%%%%%%%%%%%%%


\begin{thebibliography}{99}

\bibitem{uc}
M.~Greiner, O.~Mandel, T.~W.~H\"ansch, and I.~Bloch,
%Collapse and Revival of the Matter Wave Field of a Bose-Einstein Condensate,
Nature {\bf 419} 51 (2002).



\bibitem{kww-06}
T. Kinoshita, T. Wenger,  D. S. Weiss, %A quantum Newton's cradle,
 Nature {\bf 440}, 900 (2006).

\bibitem{tc-07}
S. Hofferberth, I. Lesanovsky, B. Fischer, T. Schumm, and J. Schmiedmayer,
%Non-equilibrium coherence dynamics in one-dimensional Bose gases,
Nature {\bf 449}, 324 (2007).

\bibitem{tetal-11}
S. Trotzky Y.-A. Chen, A. Flesch, I. P. McCulloch, U. Schollw\"ock,
J. Eisert, and I. Bloch, 
%Probing the relaxation towards equilibrium in an isolated strongly correlated 1D Bose gas,  
Nature Phys. {\bf 8}, 325 (2012). 

\bibitem{cetal-12}
M. Cheneau, P. Barmettler, D. Poletti, M. Endres, P. Schauss, T. Fukuhara, C. Gross, I. Bloch, C. Kollath, and S. Kuhr,
%Light-cone-like spreading of correlations in a quantum many-body system,
Nature {\bf 481}, 484 (2012).

\bibitem{getal-11}
M. Gring, M. Kuhnert, T. Langen, T. Kitagawa, B. Rauer, M. Schreitl, I. Mazets, D. A. Smith, E. Demler, and J. Schmiedmayer,
%Relaxation Dynamics and Pre-thermalization in an Isolated Quantum System,
Science {\bf 337}, 1318 (2012).


\bibitem{shr-12}
U. Schneider, L. Hackerm\"uller, J. P. Ronzheimer, S. Will, S. Braun, T. Best, I. Bloch, E. Demler, S. Mandt, D. Rasch, and A. Rosch,
Nature Phys. {\bf 8}, 213 (2012).


\bibitem{rsb-13}
J. P. Ronzheimer, M. Schreiber, S. Braun, S. S. Hodgman, S. Langer, I. P. McCulloch, F. Heidrich-Meisner, I. Bloch, and U. Schneider,
Phys. Rev. Lett. {\bf 110}, 205301 (2013).


\bibitem{revq}
A. Polkovnikov, K. Sengupta, A. Silva, and M. Vengalattore, %Nonequilibrium dynamics of closed interacting quantum systems, 
Rev. Mod. Phys. {\bf 83}, 863 (2011).

\bibitem{gg} 
M. Rigol, V. Dunjko, V. Yurovsky,  and M. Olshanii,
%Relaxation in a completely integrable many-body quantum system: An ab initio
%study of the dynamics of the highly excited states of lattice hard-core bosons,
Phys. Rev. Lett. {\bf 98}, 50405 (2007).

%%%%%%%% numerics, bad


\bibitem{kla-07}
C. Kollath, A. L\"auchli, and E. Altman, Phys. Rev. Lett. {\bf 98}, 180601 (2007). 

\bibitem{bch-11}
M. C. Banuls, J. I. Cirac, and M. B. Hastings,
Phys. Rev. Lett. {\bf 106}, 050405 (2011). 

\bibitem{gme-11}
C. Gogolin, M. P. Mueller, and J. Eisert, Phys. Rev. Lett. {\bf 106}, 040401 (2011).

\bibitem{gm-11} 
P. Grisins and I. E. Mazets,  Phys. Rev. A {\bf 84}, 053635 (2011).

\bibitem{gp-08}
D.M. Gangardt and M. Pustilnik, Phys. Rev. A {\bf 77}, 041604 (2008).




%%%% numerics good

\bibitem{mwn-07}
S.R. Manmana, S. Wessel, R.M. Noack, and A. Muramatsu, Phys. Rev. Lett. {\bf 98}, 210405 (2007).

\bibitem{rdo-08}
M. Rigol, V. Dunjko,  and M. Olshanii,
%Thermalization and its mechanism for generic isolated quantum systems,
Nature {\bf 452}, 854 (2008). 


\bibitem{r-09}
M. Rigol, Phys. Rev. Lett. {\bf 103}, 100403 (2009); Phys. Rev. A {\bf 80}, 053607 (2009).

\bibitem{rs-12}
M. Rigol and M. Srednicki, Phys. Rev. Lett. {\bf 108}, 110601 (2012).

\bibitem{bkl-10}
%Effect of Rare Fluctuations on the Thermalization of Isolated Quantum Systems
G. Biroli, C. Kollath, and A. Laeuchli,
Phys. Rev. Lett. {\bf 105}, 250401 (2010).

\bibitem{rsm-10}
D. Rossini, A. Silva, G. Mussardo, and G. Santoro, Phys. Rev. Lett. {\bf 102}, 127204 (2009); 
D. Rossini, S. Suzuki, G. Mussardo, G. E. Santoro, and A. Silva, Phys. Rev. B {\bf 82}, 144302 (2010).

\bibitem{gcg-11} 
L. Foini, L. F. Cugliandolo, and A. Gambassi, Phys. Rev. B {\bf 84}, 212404 (2011); 
%L. Foini, L. F. Cugliandolo, and A. Gambassi, 
J. Stat. Mech. (2012) P09011.

\bibitem{rf-11}
M. Rigol and M. Fitzpatrick, Phys. Rev. A {\bf 84}, 033640 (2011);
K. He and M. Rigol, Phys. Rev. A {\bf 85}, 063609 (2012).

\bibitem{bdkm-12}
G. P. Brandino, A. De Luca, R.M. Konik, and G. Mussardo, %Quench Dynamics in Randomly Generated Extended Quantum Models,
Phys. Rev. B {\bf 85}, 214435 (2012).

\bibitem{krs-12}
C. Karrasch, J. Rentrop, D. Schuricht, and V. Meden, Phys. Rev. Lett. {\bf 109}, 126406 (2012); 
J. Rentrop, D. Schuricht, and V. Meden, New J. Phys. {\bf 14}, 075001 (2012).

\bibitem{sks-13}
J. Sirker, N. P. Konstantinidis,  F. Andraschko, and N. Sedlmayr,
arXiv:1303.3064.


%%% attempts


\bibitem{cro}
H. Buljan, R. Pezer, and T. Gasenzer, Phys. Rev. Lett. {\bf 100}, 080406 (2008).

\bibitem{fcc-09}
A. Faribault, P. Calabrese, and J.-S. Caux, J. Stat. Mech. P03018 (2009);
%A. Faribault, P. Calabrese, and J.-S. Caux, 
J. Math. Phys. {\bf 50}, 095212 (2009).

\bibitem{grd-10}
V. Gritsev, T. Rostunov, and E. Demler, J. Stat. Mech. (2010) P05012.

\bibitem{fm-10}
D. Fioretto and G. Mussardo,
%Quantum Quenches in Integrable Field Theories,
New J. Phys. {\bf 12}, 055015 (2010).

\bibitem{sfm-12}
S. Sotiriadis, D. Fioretto, and G. Mussardo,
%Zamolodchikov-Faddeev Algebra and Quantum Quenches in Integrable Field Theories,
J. Stat. Mech. (2012) P02017.

\bibitem{mc-12}
J. Mossel and J.-S. Caux, New J. Phys. {\bf 14},  075006 (2012).


\bibitem{ck-12}
%Constructing the Generalized Gibbs Ensemble after a Quantum Quench
J.-S. Caux and R. M. Konik, Phys. Rev. Lett. {\bf 109}, 175301 (2012).

\bibitem{a-12}
D. Iyer and N. Andrei, Phys. Rev. Lett. {\bf 109}, 115304 (2012);
D. Iyer, H. Guan, and N. Andrei, Phys. Rev. A {\bf 87}, 053628 (2013).

\bibitem{ce-13}
J.-S. Caux and F. H. L. Essler, Phys. Rev. Lett. {\bf 110}, 257203 (2013).


\bibitem{bck-13}
%Relaxation dynamics of conserved quantities in a weakly non-integrable one-dimensional Bose gas
G. Brandino, J.-S. Caux, and R. M. Konik,  arXiv:1301.0308.
 



%%%free particles

\bibitem{cc-06} P. Calabrese and  J. Cardy, %Time-dependence of correlation functions following a quantum quench,
Phys. Rev. Lett. {\bf 96}, 136801 (2006); 
% P. Calabrese and  J. Cardy,  %Quantum quenches in extended systems, 
J. Stat. Mech. P06008  (2007); J. Stat. Mech. P04010 (2005).

\bibitem{c-06}
M. A. Cazalilla, Phys. Rev. Lett. {\bf 97}, 156403 (2006); 
A. Iucci, and M. A. Cazalilla, Phys. Rev. A {\bf 80}, 063619 (2009); 
New J. Phys. {\bf 12}, 055019 (2010);
A. Mitra and T. Giamarchi, Phys. Rev. Lett. {\bf 107}, 150602 (2011);
M. A. Cazalilla, A. Iucci, and M.-C. Chung, Phys. Rev. E {\bf 85}, 011133 (2012).


\bibitem{cdeo-08}
M. Cramer, C. M. Dawson, J. Eisert, and T. J. Osborne, 
%Quenching, relaxation, and a central limit theorem for quantum lattice systems, 
Phys. Rev. Lett. {\bf 100}, 030602 (2008);
M. Cramer and J. Eisert,
%A quantum central limit theorem for non-equilibrium systems: Exact local relaxation of correlated states,
New J. Phys. 12, 055020 (2010).

\bibitem{bs-08}
T. Barthel and U. Schollw\"ock, %Dephasing and the steady state in quantum many-particle systems,  
Phys. Rev. Lett. {\bf 100}, 100601 (2008).

\bibitem{CEFII}
P. Calabrese, F. H. L. Essler, and M. Fagotti, 
J. Stat. Mech. (2012) P07022.

\bibitem{scc-09}
S. Sotiriadis, P. Calabrese,  and J. Cardy, %Quantum Quench from a Thermal Initial State, 
EPL {\bf 87}, 20002, (2009).

\bibitem{CEF}
P. Calabrese, F. H. L. Essler, and M. Fagotti, 
%Quantum Quench in the Transverse Field Ising Chain,
Phys. Rev. Lett. {\bf 106}, 227203 (2011);  J. Stat. Mech. (2012) P07016.



\bibitem{f-13}
M. Fagotti,  Phys. Rev. B {\bf 87}, 165106 (2013).

\bibitem{eef-12}
%Dynamical Correlations after a Quantum Quench
F. H. L. Essler, S. Evangelisti, and M. Fagotti,
Phys. Rev. Lett. {\bf 109}, 247206 (2012).

\bibitem{se-12}
D. Schuricht and F. H. L. Essler, J. Stat. Mech. (2012) P04017.

\bibitem{fe-13}
%Reduced Density Matrix after a Quantum Quench
M.  Fagotti and  F. H. L. Essler, Phys. Rev. B {\bf 87}, 245107 (2013).

%\bibitem{ccss-11} T. Caneva, E. Canovi, D. Rossini, G. E. Santoro, and A. Silva,  J. Stat. Mech. (2011) P07015.


\bibitem{US}
M. Collura, S. Sotiriadis, and P. Calabrese, Phys. Rev. Lett. {\bf 110}, 245301 (2013);
J. Stat. Mech. (2013) P09025.

\bibitem{fost} Some non-linear problems have been studied in the presence of inhomogeneities in 
M. S. Foster, E. A. Yuzbashyan, and B. L. Altshuler, Phys. Rev. Lett. {\bf 105}, 135701 (2010);
M. S. Foster, T. C. Berkelbach, D. R. Reichman, and E. A. Yuzbashyan, Phys. Rev. B {\bf 84}, 085146 (2011).

%========Construction GGE

\bibitem{mc-12b} 
J. Mossel and J.-S. Caux, J. Phys. A {\bf 45}, 255001 (2012);
  E. Demler and A. M. Tsvelik, Phys. Rev. B {\bf 86}, 115448 (2012). 

\bibitem{p-13}
B. Pozsgay, J. Stat. Mech. (2013) P07003.

\bibitem{fe-13b}
M.  Fagotti and  F. H. L. Essler, J. Stat. Mech. (2013) P07012.

\bibitem{ksc-13}
M. Kormos, A. Shashi, Y.-Z. Chou, J.-S. Caux, and A. Imambekov,
Phys. Rev. B {\bf 88}, 205131 (2013).  %Interaction quenches in the 1D Bose gas


\bibitem{m-13}
G. Mussardo, Phys. Rev. Lett. {\bf 111}, 100401 (2013).

 \bibitem{fcce-13}
M. Fagotti, M. Collura, F. H. L. Essler, P. Calabrese,  arXiv:1311.5216.


\bibitem{fle-10}
D. Muth, B. Schmidt, and M. Fleischhauer, New J. Phys. {\bf 12}, 083065 (2010); 
D. Muth and M. Fleischhauer, Phys. Rev. Lett. {\bf 105}, 150403 (2010).


\bibitem{LiebPR130} E. H. Lieb and W. Liniger, Phys. Rev. {\bf 130}, 1605 (1963);  
E. H. Lieb, Phys. Rev. {\bf 130}, 1616 (1963).

\bibitem{TG} L. Tonks, Phys. Rev. {\bf 50}, 955 (1936);  M. Girardeau, J. Math. Phys. {\bf 1}, 516 (1960).

%\bibitem{SM}
%See Supplemental Material for a rigorous derivation of the fermionic multi-point correlations on the lattice.

\bibitem{dk-90}
B. Davies, Physica A {\bf 167}, 433 (1990); B. Davies and V. E. Korepin, arXiv:1109.6604. 

\bibitem{nm-13}
S. S. Natu and E. J. Mueller, 	Phys. Rev. A {\bf 87}, 053607 (2013).

\bibitem{kpz}
P. Calabrese and P. Le Doussal, Phys. Rev. Lett. {\bf 106}, 250603 (2011);
P. Le Doussal and P. Calabrese, J. Stat. Mech. (2012) P06001.

\bibitem{tan}
M. Olshanii and V. Dunjko, Phys. Rev. Lett. {\bf 91},  090401 (2003);
S. Tan, Ann. Phys. {\bf 323}, 2952 (2008);
M. Barth and W. Zwerger, Ann. Phys. {\bf 326}, 2544  (2011).




\end{thebibliography}
\end{document}